\documentclass[prl,unsortedaddress,superscriptaddress]{revtex4}
\usepackage{amsmath}  
\usepackage{amsbsy,amssymb,amsfonts}
\usepackage{graphicx}
\usepackage{color}
\usepackage{epsf}
\usepackage[utf8]{inputenc}
\usepackage{hyperref}
\usepackage{multirow}
\def\cm{$\rm cm^{-1}$}

\def\bravert{\egroup\,\vrule\,\bgroup}

{\catcode`\|=\active
  \gdef\Twoint#1{\left(\mathcode`\|"8000\let|\bravert {#1}\right)}}
{\catcode`\|=\active
  \gdef\Braket#1{\left<\mathcode`\|"8000\let|\bravert {#1}\right>}}

\newcommand{\beq}{\begin{equation}}
\newcommand{\eeq}{\end{equation}}
\newcommand{\beqa}{\begin{eqnarray}}
\newcommand{\eeqa}{\end{eqnarray}}
\newcommand{\bea}{\begin{array}}
\newcommand{\eea}{\end{array}}

\newcommand{\bef}{\begin{figure}}
\newcommand{\ef}{\end{figure}}
\newcommand{\bc}{\begin{center}}
\newcommand{\ec}{\end{center}}
\newcommand{\bt}{\begin{table}}
\newcommand{\et}{\end{table}}
\newcommand{\btb}{\begin{tabular}}
\newcommand{\etb}{\end{tabular}}

\newcommand{\au}{{\em a.u.}}

\begin{document}

\vspace{2cm}
\title {{
         ${\cal{P,T}}$-Odd and Magnetic Hyperfine Interaction Constants and Excited-State Lifetime for HfF$^+$
       }}

\vspace*{2cm}

\author{Timo Fleig}
\email{timo.fleig@irsamc.ups-tlse.fr}
\affiliation{Laboratoire de Chimie et Physique Quantiques,
             IRSAMC, Universit{\'e} Paul Sabatier Toulouse III,
             118 Route de Narbonne, 
             F-31062 Toulouse, France}

\date{\today}
\vspace*{1cm}
\begin{abstract}
Parity- and time-reversal-symmetry violating interaction constants required for
the interpretation of a recent measurement (arXiv:1704.07928 [physics.atom-ph]) of corresponding 
symmetry violations in the $\Omega=1$ (${^3\Delta}_1$) science state of the HfF$^+$ molecular
ion are reported. Using a relativistic four-component all-electron multi-reference configuration interaction
model the nucleon-electron scalar-pseudoscalar interaction constant is determined as $W_S = 20.0$ 
[kHz]. An updated result for the electron electric-dipole-moment effective electric field
of $|E_{\text{eff}}| = 22.7 \left[\frac{\rm GV}{\rm cm}\right]$ is obtained. Further results
of relevance in the context of the search for leptonic charge-parity violation such as magnetic 
hyperfine interaction constants and electronic $G$-tensor for HfF$^+$ are presented.
\end{abstract}

\maketitle

\section{Introduction}
\label{SEC:INTRO}
Electric dipole moments (EDMs) are a powerful probe of physics beyond the standard model (SM)
of elementary particles in the regime of low energies \cite{EDMsNP_PospelovRitz2005}. The underlying
${\cal{CP}}$ (charge-parity)-violating interactions \cite{Kobayashi,FlavorPhysLepDipMom_EPJC2008} 
give rise to ${\cal{T}}$ (time-reversal)-odd energy shifts which can be strongly enhanced in atomic
matter including heavy nuclei \cite{sandars_PL1965,sushkov_flambaum1978}. In polar molecules sensitive
to lepton EDMs this enhancement can be several orders of magnitude larger than in corresponding
heavy atoms \cite{commins_demille_EDM_2008}. Such molecular systems thus hold great promise for 
the detection of a ${\cal{P,T}}$-odd energy shift which in turn could help unravel observations in
the universe yet unexplained by the SM, in particular related to its matter and energy content 
\cite{Sakharov_JETP1967,Hinshaw:2012aka,Dine_Kusenko_MatAntimat2004}.

HfF$^+$ is a paramagnetic system where the major sources of a potential ${\cal{P,T}}$-odd molecular 
EDM are the electron EDM, $d_e$, and the nucleon-electron scalar-pseudoscalar (ne-SPS) interaction, $C_S$
\cite{Barr_eN-EDM_Atoms_1992}. The sensitivity of HfF$^+$ to these two ${\cal{CP}}$-violating
parameters allows for interpreting a null measurement and the upper bound to the molecular EDM through
the corresponding interaction constants, $W_d$ for the electron EDM and $W_S$ for the ne-SPS
interaction, respectively. The main objective of the present work is, therefore, the accurate calculation
of $W_S$ and a reassessment of $W_d$ for the $\Omega = 1$ ({$^3\Delta_1$}) state in which a recent 
measurement yielded an EDM-sensitive frequency shift of $f^{BD} = 0.10 \pm 0.87_{\text{stat}} \pm 
0.20_{\text{sys}}$\, [mHz] \cite{Cairncross:2017fip,Cornell_MolIons_Science2013}. The measured
shift is consistent with zero and this result and the presently calculated interaction constants for
HfF$^+$ are likely \cite{Chupp_Ramsey_Global2015} to lead to stronger constraints on $d_e$ and $C_S$, 
since the tightest upper bound on an EDM shift in a paramagnetic system thus far has been obtained from 
measurements on the ThO molecule \cite{ACME_ThO_eEDM_science2014,ACME_ThO_ArXiV2016} where the heavy
atom has a significantly larger proton number.

The following section briefly reviews definitions of interaction constants, the electronic $G$ tensor,
and the method for calculating the lifetime of the relevant excited state in which the EDM experiment
has been carried out. In the final section electronic-structure methods and technical parameters are
defined, followed by a discussion of the present results and their consequences in view of further
constraining ${\cal{CP}}$-violation in the lepton sector.

\section{Theory}
\label{SEC:THEORY}
\subsection{${\cal{P,T}}$-odd interaction constants}

The electron EDM interaction constant is evaluated as proposed in stratagem II of Lindroth et al.
\cite{lindroth_EDMtheory1989} as an effective one-electron operator via the squared electronic
momentum operator,
\begin{equation}
 E_{\text{eff}} = \frac{2\imath c}{e\hbar} \left< \Psi_{\Omega} \right|
 \sum\limits_{j=1}^n\, \gamma^0_j \gamma^5_j\, \vec{p}\,^2_j \left| \Psi_{\Omega} \right>
\end{equation}
with $n$ the number of electrons and $j$ an electron index, as described in greater detail in
reference \cite{fleig_nayak_eEDM2013}.
The EDM effective electric field is related to the electron EDM interaction constant
$W_d = -\frac{1}{\Omega}\, E_{\text{eff}}$.

The ne-SPS interaction constant is defined and implemented \cite{ThF+_NJP_2015} as
\begin{equation}
 W_S = \frac{\imath}{\Omega}\,
 \frac{G_F}{\sqrt{2}}\, Z\, \left< \Psi_{\Omega} \right| \sum\limits^n_{j=1}\, {\gamma^0_j\gamma^5_j\, \rho_K(\vec{r}_j)}
     \left| \Psi_{\Omega}  \right>
\end{equation}
where $G_F$ is the Fermi constant, $Z$ is the proton number and $\rho_K(\vec{r}_j)$ is the nuclear charge
density at position $\vec{r}_j$, normalized to unity.

\subsection{Magnetic hyperfine interaction}

The parallel component of the magnetic hyperfine interaction tensor is defined as follows (\au):
\begin{equation}
A_{||}(K) = \frac{\mu_{K}[\mu_N]}{2cIm_p \Omega}\,
          \left< \Psi_{\Omega} \right| \sum\limits_{i=1}^n\, \left( \frac{\vec{\alpha_i} \times \vec{r}_{iK}}{r_{iK}^3}
          \right)_z \left| \Psi_{\Omega} \right>
\end{equation}
where $\mu_{K}[\mu_N]$ is the magnetic moment of nucleus $K$ in Bohr magnetons, $\vec{\alpha}$ is a vector of Dirac 
matrices and $\vec{r}_{iK}$ is the position vector relative to nucleus $K$. Further details can be found in reference 
\cite{Fleig2014}.

\subsection{Excited-State Lifetime}

The decay rate of an electronically excited state $\psi_k$ to a state $\psi_a$ {\it{via}} spontaneous 
emission of photons reads in electric-dipole (E1) approximation
\begin{equation}
 \label{EQ:DECAY_RATE}
 \Gamma_{ka} = \frac{4}{3}\, \frac{\left( \varepsilon_k - \varepsilon_a \right)^3}{\hbar^4 c^3}\,
                          \left| {\bf{D}}_{ak} \right|^2
\end{equation}
where $\varepsilon_k - \varepsilon_a$ is the transition energy, $D_{ak}$ is an electric transition dipole 
matrix element between states $\Psi_a$ and $\Psi_k$ and $\hat{\bf{D}}$ is the electronic electric dipole 
moment vector operator.

The ensuing lifetime $\tau$ of the state $\psi_k$ is determined as
\begin{equation}
 \label{EQ:LIFETIME}
 \tau_{k} = \Gamma_{ka}^{-1}
\end{equation}
where only decay into state $\psi_a$ (usually the electronic ground state) has been taken into account.

\subsection{Electronic G-Tensor}

The parallel component of the electronic G-tensor for a linear molecule is defined as
\begin{equation}
 \label{EQ:GTENSOR_PAR}
 G_{||} = \frac{1}{\Omega}\, 
          \left< \psi_{\Omega} | \hat{L}^e_{\hat{n}} + g_s \hat{S}^e_{\hat{n}} | \psi_{\Omega} \right>
 \end{equation}
with $\hat{L}^e_{\hat{n}} = \hat{\bf{L}}^e \cdot \hat{n}$ where $\hat{n}$ is a unit vector along the 
molecular axis and $g_s = -g_e = 2.00231930436182$ is the free-electron g-factor \cite{nist_g-2}.

Thus, for a molecule in a ${^3\Delta}_1$ state $\Omega = 1$, $\Lambda = \pm 2$, and 
$\Sigma = \mp 1$, so $G_{||}({^3\Delta}_1) \approx 0$. However, spin-orbit interaction will
mix states with deviant $\Lambda$ and $\Sigma$ quantum numbers into the
state denoted ${^3\Delta}_1$, {\it{e.g.}} a higher-lying ${^1\Pi}_1$ state {\it{via}}
\begin{equation}
 \left< {^3\Delta}_1 | \hat{H}_{SO} | {^1\Pi}_1 \right> \neq 0
\end{equation}
where $\hat{H}_{SO}$ is a generic spin-orbit interaction Hamiltonian.
It is, therefore, of importance to numerically determine $G_{||}({^3\Delta}_1)$ using molecular electronic
wavefunctions including the effects of spin-orbit interaction.

\section{Application to HfF$^+$}
\label{SEC:APPL}
\subsection{Technical details}

Atomic basis sets are of triple-$\zeta$ quality \cite{dyall_basis_2004,dyall_gomes_basis_2010,Dunning_jcp_1989}
including core- and valence-correlating functions. The molecular spinor basis is obtained in Hartree-Fock
approximation using the Dirac-Coulomb Hamiltonian (here in \au)
\begin{equation}
  \hat{H}^{DC} = \sum\limits^n_i\, \left[ c\, \boldsymbol{\alpha}_i \cdot {\bf{p}}_i
          + \beta_i c^2 + \sum\limits^N_A\, \frac{Z_A}{r_{iA}}{1\!\!1}_4 \right] + \sum\limits^n_{i,j>i}\, 
                                             \frac{1}{r_{ij}}{1\!\!1}_4 + V_{AB}
\end{equation}
where $\alpha$ and $\beta$ are Dirac matrices, $Z_A$ is the proton number for nucleus $A$, $i,j$ are electron
indices and $V_{AB}$ is the classical electrostatic potential energy for the clamped nuclei. For the Fock operator
in the Dirac-Coulomb-Hartree-Fock calculation valence configurations are averaged over Hf($6s^2$), 
Hf($6s^1 5d_{\delta}^1$), and Hf($5d_{\delta}^2$). For correlated calculations the wavefunction model
MR$_{12}$-CISD(20) is used in the present. It corresponds to the parameter settings for the model
MR-CISD(20) in ref. \cite{fleig_nayak_eEDM2013}, however, using a larger active spinor space by adding 
four $\pi$-type and two $\sigma$-type Kramers pairs (all of Hf atomic character) from the virtual into the next 
lower space. Since for the eEDM effective electric field it has been shown that the inclusion of higher excitation 
ranks and the correlation of Hf outer-core electrons leads to only minor corrections which even largely cancel
each other, the current model with $20$ correlated electrons is considered sufficiently accurate.
Further technical details on molecular wavefunctions can be found in reference \cite{fleig_nayak_eEDM2013}. 

For the determination of parallel magnetic hyperfine interaction constants ($A_{||}$) the following nuclear
magnetic moments have been used \cite{stone_INDC2015}: $\mu = 0.7936\, \mu_N$ for {$^{177}$Hf} with nuclear spin 
$I = 7/2$ and $\mu = 2.62887\, \mu_N$ for {$^{19}$F} with nuclear spin $I = 1/2$.

\subsection{Results and discussion}

All calculated constants for HfF$^+$ are given in Table \ref{TAB:HFF+}. 

The g-tensor component $G_{||}(\Omega = 1) \approx 0.013$ \au\ is small but still roughly an order of magnitude larger
than $g_s-2 \approx 0.00232$. Since spin-orbit interaction is treated rigorously in the present approach,
neither $\Lambda$ nor $\Sigma$ (or $S$, for that matter) are exact quantum numbers of the CI state. Therefore,
the state denoted $\Omega = 1$ is a mixture of terms ${^3\Delta}_1$, ${^1\Pi}_1$, and others. The determined
value for $G_{||}(\Omega = 1)$ indicates that a small contribution to the science state from a low-lying 
${^1\Pi}_1$ state, arising from an excitation $6s^1 \rightarrow 6p^1$ on Hf is likely to be the main source of
``contamination'', since $G_{||}({^1\Pi}_1) = +1$.

For the calculation of the excited-state lifetime from the decay rate (Eq. (\ref{EQ:DECAY_RATE})) into the 
electronic ground state ($\Omega=0$) the experimental excitation energy of $991.83$ \cm\ 
\cite{Cossel_HfF+_CPL2012} has been used. The required electric transition-dipole matrix element is
with the model vTZ/MR$_{12}$-CISD(20) calculated as
$\left|\left|\left< \psi_{\Omega=1} | {\hat{\vec{D}}} | \psi_{\Omega=0} \right>\right|\right| \approx 0.0349$ 
[Debye], yielding a lifetime of $\tau_{\Omega=1} \approx 2.7$ [s] which agrees well with the measured lifetime of 
$\tau_{\text{exp}} = 2.1(1)$ [s] \cite{Cairncross:2017fip}.

The magnetic hyperfine interaction constant for the {$^{19}$F} nucleus agrees qualitatively with the value of
$-62.0(2)$ [MHz] measured in reference \cite{Cairncross:2017fip}. However, the present electronic-structure model
has not been designed with a focus on properties depending on spin density in the vicinity of the fluorine nucleus.
On the other hand, the calculated hyperfine interaction constant for the {$^{177}$Hf} nucleus allows for an
assessment of the accuracy of the molecular wavefunction for the electron EDM and ne-SPS interaction constants.
At present, the author is not aware of a measurement of $A_{||}$({$^{177}$Hf}) in the $\Omega=1$ state of HfF$^+$
to compare with.

The electron EDM effective electric
field has been recalculated with the present slightly improved correlation model, and a small decrease of
its value by $-0.6\, \left[\frac{\rm GV}{\rm cm}\right]$ (or $2.5$\%) relative to the result reported in
reference \cite{fleig_nayak_eEDM2013} is observed. A related drop of $E_{\text{eff}}$ also occurs in the 
valence isoelectronic ThF$^+$ molecular ion \cite{ThF+_NJP_2015} and its origin is explained in reference 
\cite{Fleig2014}.

${\cal{P,T}}$-odd energy shifts are defined as $\Delta E_{\cal{P,T}} = -d_e\, E_{\text{eff}} = d_e\, W_d\, \Omega$ 
for the electron EDM,
where $W_d = -E_{\text{eff}}/\Omega$, and $\Delta E_{\cal{P,T}} = C_S\, W_S\, \Omega$
for the ne-SPS interaction.
Therefore, the ratio of the energy shifts due to leptonic ($W_d$) and semi-leptonic ($W_S$) interaction 
constants is proportional to the ratio of these interaction constants. Thus,
\begin{equation}
 \frac{W_d}{W_S} = \frac{-E_{\text{eff}}}{\Omega\, W_S}.
 \label{EQ:WDWSRATIO}
\end{equation}
With the interaction constants determined in the present work, 
$\frac{W_d}{W_S}$(HfF$^+$) $\approx 274.5 \left[ \frac{10^{18}}{e\, {\rm{cm}}} \right]$ 
for the science state with $\Omega = 1$. This ratio differs significantly from the one for ThO ($\Omega = 1$)
where $\frac{W_d}{W_S} \approx 172 \left[ \frac{10^{18}}{e\, {\rm{cm}}} \right]$ \cite{Denis-Fleig_ThO_JCP2016}
which is predominantly due to the large difference in proton number between Th ($Z=92$) and Hf ($Z=72$) 
\cite{PhysRevA.84.052108,PhysRevA.85.029901}. Consequently, accounting for both leading ${\cal{P,T}}$-odd
enhancements in the paramagnetic system HfF$^+$ and the result of the recent measurement \cite{Cairncross:2017fip}
will likely lead to stronger constraints on the ${\cal{CP}}$-violating parameters $d_e$ and $C_S$ through
the procedure of global fits \cite{MJung_robustlimit2013,Chupp_Ramsey_Global2015}.

The more detailed discussion of electronic-structure models for HfF$^+$ in reference \cite{fleig_nayak_eEDM2013}
suggests that the uncertainty for both $W_d$ and $W_S$ presented here remains at $5$\%, although a more accurate
active-space model has been employed in the present work. The same applies to the heavy-atom magnetic hyperfine
constant $A_{||}$({$^{177}$Hf}) with reference to earlier studies of hyperfine interaction on valence isoelectronic
systems \cite{ThF+_NJP_2015,Denis-Fleig_ThO_JCP2016}. The uncertainties for $G_{||}$ and the molecule-frame
EDM $D$ are estimated to be $10$\% and $15$\%, respectively, in accord with the discussion in reference 
\cite{PhysRevA.95.022504}.


\newpage

\section{Tables}

\begin{table}[h]

\caption{\label{TAB:HFF+}
         Molecule-frame static electric dipole moment,
         electron EDM and ne-S-PS interaction constants, parallel component of electronic G tensor, 
	 excited-state lifetime, and magnetic hyperfine
         interaction constants for the $\Omega = 1$ (${^3\Delta}_1$) science state of HfF$^+$.
	 All values are given for the Hf nucleus at the origin of the reference frame and the F
	 nucleus at $-3.4384$ \au\ on the $z$ axis. The molecule-frame dipole moment corresponds
	 to an origin located at the center of mass.
        }

 \vspace*{0.5cm}
 \hspace*{-1.7cm}
 \begin{center}
  \begin{tabular}{c|ccccccc}
     CI Wavefunction model   & $D$ [Debye]
                             & $E_{\text{eff}} \left[\frac{\rm GV}{\rm cm}\right]$ 
                             & $W_S$ [kHz]
                             & $G_{||}$ [a.u.]
                             & $\tau$ [s]  
			     & $A_{||}$({$^{177}$Hf}) [MHz]
			     & $A_{||}$({$^{19}$F}) [MHz]
  \vspace*{0.05cm} \\ \hline
   vTZ/MR$_{12}$-CISD(20)    & $4.19$  & $-22.7$  &  $20.0$  &  $0.0127$  &  $2.7$  & $-1328$  & $-43.0$ 
  \end{tabular}
 \end{center}
\end{table}


\newpage

\bibliographystyle{unsrt}

\begin{thebibliography}{10}

\bibitem{EDMsNP_PospelovRitz2005}
M.~Pospelov and A.~Ritz.
\newblock Electric dipole moments as probes of new physics.
\newblock {\em Ann. Phys.}, 318:119, 2005.

\bibitem{Kobayashi}
M.~Kobayashi and T.~Maskawa.
\newblock {CP violation in the renormalizable theory of weak interaction}.
\newblock {\em Prog. Theor. Phys.}, 49:652, 1973.

\bibitem{FlavorPhysLepDipMom_EPJC2008}
M.~Raidal, A.~van~der Schaaf, I.~Bigi, M.~L. Mangano, Y.~Semertzidis, S.~Abel,
  S.~Albino, S.~Antusch, E.~Arganda, B.~Bajc, S.~Banerjee, C.~Biggio,
  M.~Blanke, W.~Bonivento, G.~C. Branco, D.~Bryman, A.~J. Buras, L.~Calibbi,
  A.~Ceccucci, P.~H. Chankowski, S.~Davidson, A.~Deandrea, D.~P. DeMille,
  F.~Deppisch, M.~A. Diaz, B.~Duling, M.~Felcini, W.~Fetscher, F.~Forti, D.~K.
  Ghosh, M.~Giffels, M.~A. Giorgi, G.~Giudice, E.~Goudzovskij, T.~Han, P.~G.
  Harris, M.~J. Herrero, J.~Hisano, R.~J. Holt, K.~Huitu, A.~Ibarra,
  O.~Igonkina, A.~Ilakovac, J.~Imazato, G.~Isidori, F.~R. Joaquim, M.~Kadastik,
  Y.~Kajiyama, S.~F. King, K.~Kirch, M.~G. Kozlov, M.~Krawczyk, T.~Kress,
  O.~Lebedev, A.~Lusiani, E.~Ma, G.~Marchiori, A.~Masiero, L.~Masina,
  G.~Moreau, T.~Mori, M.~Muntel, N.~Neri, F.~Nesti, C.~G.~J. Onderwater,
  P.~Paradisi, S.~T. Petcov, M.~Picariello, V.~Porretti, A.~Poschenrieder,
  M.~Pospelov, L.~Rebane, M.~N. Rebelo, A.~Ritz, L.~Roberts, A.~Romanino, J.~M.
  Roney, A.~Rossi, R.~R{\"u}ckl, G.~Senjanovic, N.~Serra, T.~Shindou,
  Y.~Takanishi, C.~Tarantino, A.~M. Teixeira, E.~Torrente-Lujan, K.~J.
  Turzynski, T.~E.~J. Underwood, S.~K. Vempati, and O.~Vives.
\newblock Flavor physics of leptons and dipole moments.
\newblock {\em Eur. Phys. J. C}, 57:13, 2008.

\bibitem{sandars_PL1965}
P.~G.~H. Sandars.
\newblock The electric dipole moment of an atom.
\newblock {\em Phys. Lett.}, 14:194, 1965.

\bibitem{sushkov_flambaum1978}
O.~P. Sushkov and V.~V. Flambaum.
\newblock Parity breaking effects in diatomic molecules.
\newblock {\em Sov. Phys. JETP}, 48:608, 1978.

\bibitem{commins_demille_EDM_2008}
B.~L. Roberts and W.~J. Marciano, editors.
\newblock {\em Lepton {D}ipole {M}oments}, volume~20, chapter~14, pages 519 --
  581.
\newblock Advances Series on Directions in High Energy Physics, World
  Scientific, New Jersey, 2009.
\newblock E. D. Commins and D. DeMille, The {E}lectric {D}ipole {M}oment of the
  {E}lectron.

\bibitem{Sakharov_JETP1967}
A.~D. Sakharov.
\newblock Violation of {CP} invariance, {C} asymmetry, and baryon asymmetry of
  the universe.
\newblock {\em {JETP} Lett.}, 5:24, 1967.

\bibitem{Hinshaw:2012aka}
G.~Hinshaw et~al.
\newblock {Nine-Year Wilkinson Microwave Anisotropy Probe (WMAP) Observations:
  Cosmological Parameter Results}.
\newblock {\em Astrophys. J. Suppl.}, 208:19, 2013.

\bibitem{Dine_Kusenko_MatAntimat2004}
M.~Dine and A.~Kusenko.
\newblock The origin of the matter-antimatter asymmetry.
\newblock {\em Rev. Mod. Phys.}, 76:1, 2004.

\bibitem{Barr_eN-EDM_Atoms_1992}
S.~M. Barr.
\newblock ${T}$- and ${P}$-odd electron-nucleon interactions and the electric
  dipole moments of large atoms.
\newblock {\em Phys. Rev. D}, 45:4148, 1992.

\bibitem{Cairncross:2017fip}
William~B. Cairncross, Daniel~N. Gresh, Matt Grau, Kevin~C. Cossel, Tanya~S.
  Roussy, Yiqi Ni, Yan Zhou, Jun Ye, and Eric~A. Cornell.
\newblock {A precision measurement of the electron's electric dipole moment
  using trapped molecular ions}.
\newblock 2017.
\newblock arXiv:1704.07928 [physics.atom-ph].

\bibitem{Cornell_MolIons_Science2013}
H.~Loh, K.~Cossel, M.~C. Grau, K.-K. Ni, E.~R. Meyer, J.~L. Bohn, J.~Ye, and
  E.~A. Cornell.
\newblock Precision {S}pectroscopy of {P}olarized {M}olecules in an {I}on
  {T}rap.
\newblock {\em Science}, 342:1220, 2013.

\bibitem{Chupp_Ramsey_Global2015}
T.~Chupp and M.~Ramsey-Musolf.
\newblock {E}lectric dipole moments: {A} global analysis.
\newblock {\em Phys. Rev. C}, 91:035502, 2015.

\bibitem{ACME_ThO_eEDM_science2014}
The~{ACME} Collaboration, J.~Baron, W.~C. Campbell, D.~DeMille, J.~M. Doyle,
  G.~Gabrielse, Y.~V. Gurevich, P.~W. Hess, N.~R. Hutzler, E.~Kirilov,
  I.~Kozyryev, B.~R. O'Leary, C.~D. Panda, M.~F. Parsons, E.~S. Petrik,
  B.~Spaun, A.~C. Vutha, and A.~D. West.
\newblock Order of {M}agnitude {S}maller {L}imit on the {E}lectric {D}ipole
  {M}oment of the {E}lectron.
\newblock {\em Science}, 343:269, 2014.

\bibitem{ACME_ThO_ArXiV2016}
{ACME} {C}ollaboration: J~Baron, W~C Campbell, D~DeMille, J~M Doyle,
  G~Gabrielse, Y~V Gurevich, P~W Hess, N~R Hutzler, E~Kirilov, I~Kozyryev, B~R
  O'Leary, C~D Panda, M~F Parsons, B~Spaun, A~C Vutha, A~D West, and E~P West.
\newblock {M}ethods, {A}nalysis, and the {T}reatment of {S}ystematic {E}rrors
  for the {E}lectron {E}lectric {D}ipole {M}oment {S}earch in {T}horium
  {M}onoxide.
\newblock arXiv:1612.09318v1 [physics.atom-ph], 2016.

\bibitem{lindroth_EDMtheory1989}
E.~Lindroth, B.~W. Lynn, and P.~G.~H. Sandars.
\newblock Order $\alpha^2$ theory of the atomic electric dipole moment due to
  an electric dipole moment on the electron.
\newblock {\em J. Phys. B}, 22:559, 1989.

\bibitem{fleig_nayak_eEDM2013}
T.~Fleig and M.~K. Nayak.
\newblock Electron electric-dipole-moment interaction constant for {H}f{F}$^+$
  from relativistic correlated all-electron theory.
\newblock {\em Phys. Rev. A}, 88:032514, 2013.

\bibitem{ThF+_NJP_2015}
M.~Denis, M.~N{\o}rby, H.~J.~{\Aa}. Jensen, A.~S.~P. Gomes, M.~K. Nayak,
  S.~Knecht, and T.~Fleig.
\newblock Theoretical study on {T}h{F}$^+$, a prospective system in search of
  time-reversal violation.
\newblock {\em New J. Phys.}, 17:043005, 2015.

\bibitem{Fleig2014}
T.~Fleig and M.~K. Nayak.
\newblock Electron electric dipole moment and hyperfine interaction constants
  for {T}h{O}.
\newblock {\em J. Mol. Spectrosc.}, 300:16, 2014.

\bibitem{nist_g-2}
P.~J. Mohr, D.~B. Newell, and B.~N. Taylor.
\newblock Codata recommended values of the fundamental physical constants:
  2014.
\newblock arXiv:1507.07956v1 [physics.atom-ph], 2014.

\bibitem{dyall_basis_2004}
K.~G. Dyall.
\newblock Relativistic double-zeta, triple-zeta, and quadruple-zeta basis sets
  for the 5d elements {H}f-{H}g.
\newblock {\em Theoret. Chim. Acta}, 112:403, 2004.

\bibitem{dyall_gomes_basis_2010}
K.~G. Dyall and A.~S.~P. Gomes.
\newblock Revised relativistic basis sets for the 5d elements {H}f-{H}g.
\newblock {\em Theoret. Chim. Acta}, 125:97, 2010.

\bibitem{Dunning_jcp_1989}
T.~H.~Jr. Dunning.
\newblock Gaussian basis sets for use in correlated molecular calculations.
  {I}. the atoms boron through neon and hydrogen.
\newblock {\em J. Chem. Phys.}, 90:1007, 1989.

\bibitem{stone_INDC2015}
N.~J. Stone.
\newblock {\em TABLE OF NUCLEAR MAGNETIC DIPOLE AND ELECTRIC QUADRUPOLE
  MOMENTS}.
\newblock {IAEA} {N}uclear {D}ata {S}ection {V}ienna {I}nternational {C}entre,
  {P.O.} Box 100, 1400 {V}ienna, {A}ustria, 2014.
\newblock {INDC} {I}nternational {N}uclear {D}ata {C}ommittee.

\bibitem{Cossel_HfF+_CPL2012}
K.~C. Cossel, D.~N. Gresh, L.~C. Sinclair, T.~Coffrey, L.~V. Skripnikov, A.~N.
  Petrov, N.~S. Mosyagin, A.~V. Titov, R.~W. Field, E.~R. Meyer, E.~A. Cornell,
  and J.~Ye.
\newblock Broadband velocity modulation spectroscopy of {H}f{F}$^+$: {T}owards
  a measurement of the electron electric dipole moment.
\newblock {\em Chem. Phys. Lett.}, 546:1, 2012.

\bibitem{Denis-Fleig_ThO_JCP2016}
M.~Denis and T.~Fleig.
\newblock In search of discrete symmetry violations beyond the standard model:
  {T}horium monoxide reloaded.
\newblock {\em J. Chem. Phys.}, 145:028645, 2016.

\bibitem{PhysRevA.84.052108}
V.~A. Dzuba, V.~V. Flambaum, and C.~Harabati.
\newblock Relations between matrix elements of different weak interactions and
  interpretation of the parity-nonconserving and electron
  electric-dipole-moment measurements in atoms and molecules.
\newblock {\em Phys. Rev. A}, 84:052108, Nov 2011.

\bibitem{PhysRevA.85.029901}
V.~A. Dzuba, V.~V. Flambaum, and C.~Harabati.
\newblock Relations between matrix elements of different weak interactions and
  interpretation of the parity-nonconserving and electron
  electric-dipole-moment measurements in atoms and molecules.
\newblock {\em Phys. Rev. A}, 84:052108, 2011.
\newblock Erratum ibid, {\bf 85}, 029901 (2012).

\bibitem{MJung_robustlimit2013}
M.~Jung.
\newblock {A} robust limit for the electric dipole moment of the electron.
\newblock {\em J. High Energy Phys.}, 5:168, 2013.

\bibitem{PhysRevA.95.022504}
Timo Fleig.
\newblock {T}a{O}$^{+}$ as a candidate molecular ion for searches of physics
  beyond the standard model.
\newblock {\em Phys. Rev. A}, 95:022504, Feb 2017.

\end{thebibliography}
\newcommand{\Aa}[0]{Aa}

\end{document}